\documentclass{aa} 

\usepackage{graphicx}
\usepackage{url}
\usepackage[utf8]{inputenc}
\usepackage{txfonts}
\usepackage{longtable}
\usepackage{hyperref}
\usepackage{natbib}

\begin{document}

   \title{Evidence for ram pressure stripping in a cluster of galaxies at $z$ = 0.7 \thanks{Based on observations made with ESO telescopes at the Paranal Observatory under programs 097.A-0254, 100.A-0607, 101.A-0282.}
   }
   \subtitle{}
  \author{A. Boselli\inst{1}
    \and
        B. Epinat\inst{2, 1}
    \and
        T. Contini\inst{2}
    \and
        V. Abril-Melgarejo\inst{1}
    \and
        L. A. Boogaard\inst{3}
    \and
        E. Pointecouteau \inst{2}
    \and
        E. Ventou \inst{2}
    \and
        J. Brinchmann\inst{3, 4}
    \and
        D. Carton\inst{5}
    \and
        H. Finley \inst{6,2}
    \and
        L. Michel-Dansac\inst{5}
    \and
        G. Soucail\inst{2}
    \and
        P. M. Weilbacher\inst{7}
       }

\institute{
            Aix Marseille Univ, CNRS, CNES, LAM, Marseille, France
            \email{alessandro.boselli@lam.fr, benoit.epinat@lam.fr}
        \and
            IRAP, Université de Toulouse, CNRS, CNES, UPS, (Toulouse), France
        \and
            Leiden Observatory, Leiden University, P.O. Box 9513, 2300 RA Leiden, The Netherlands
        \and
            Instituto de Astrof{\'\i}sica e Ci{\^e}ncias do Espaço, Universidade do Porto, CAUP, Rua das Estrelas, PT4150-762 Porto, Portugal
        \and
            Univ Lyon, Univ Lyon1, Ens de Lyon, CNRS, Centre de Recherche Astrophysique de Lyon UMR5574, F-69230, Saint-Genis-Laval, France
        \and
            Stockholm University, Department of Astronomy and Oskar Klein Centre for Cosmoparticle Physics, AlbaNova, University Centre SE-10691, Stockholm, Sweden
        \and
            Leibniz-Institut für Astrophysik Potsdam (AIP), An der Sternwarte 16, D-14482 Potsdam, Germany
            }

\titlerunning{Evidence for ram pressure stripping in a cluster of galaxies at $z$ = 0.73}

   \date{}

 
  \abstract
{MUSE observations of the cluster of galaxies CGr32 ($M_{200}$ $\simeq$ 2 $\times$ 10$^{14}$ M$_{\sun}$) at $z$ = 0.73 
reveal the presence of two massive star forming galaxies with extended 
tails of diffuse gas detected in the [\ion{O}{ii}]$\lambda\lambda$3727-3729\AA \ emission-line doublet. The tails, which have a cometary shape with a typical surface brightness 
of a few 10$^{-18}$ erg s$^{-1}$ cm$^{-2}$ arcsec$^{-2}$, extend up to $\simeq$ 100 kpc (projected distance) from the galaxy discs and are not associated to 
any stellar component. All this observational evidence suggests that the gas has been removed during a ram-pressure stripping event. 
This observation is thus the first evidence that dynamical interactions with the intracluster medium were active when the Universe had only
half of its present age. The density of the gas derived using the observed [\ion{O}{ii}]$\lambda$3729/[\ion{O}{ii}]$\lambda$3726 line ratio 
implies a very short recombination time, suggesting that a source of ionisation is necessary to keep the gas ionised within the tail.
}

   \keywords{Galaxies: clusters: general ; Galaxies: clusters: individual: CGr32; Galaxies: evolution; Galaxies: interactions; Galaxies: ISM; Galaxies: high-redshift
               }

   \maketitle
%

\section{Introduction}

Since the seminal work of \citet{Dressler80} it became evident that the environment plays a major role in shaping galaxy evolution.
High density regions such as rich clusters of galaxies are mainly composed of early-type objects (ellipticals and lenticulars)
while the field is dominated by late-type, star-forming systems. In these dense regions the fraction of quiescent galaxies 
increases with decreasing redshift \citep[e.g.][]{Dressler+97}. 
It is still unclear which, among the different mechanisms proposed in the literature - gravitational interactions \citep{Merritt83,
Byrd+90, Moore+98} or hydrodynamic interactions with the hot ($T$ $\simeq$ 10$^7$-10$^8$ K)
and dense ($\rho_\mathrm{ICM}$ $\simeq$ 10$^{-3}$ cm$^{-3}$, e.g. \citealp{Sarazin86}) intracluster medium \citep[ICM;][]{Gunn+72, Cowie+77, 
Nulsen82, Larson+80} - whose contribution might change at different cosmic epochs, 
is at the origin of these observed differences \citep{Boselli+06, Boselli+14}.

In the local Universe, where high sensitivity multi-frequency observations are available, allowing the
detection of the different stellar components and gas phases (cold, ionised, hot) with a
spectacular angular resolution, it is becoming evident that within structures as massive as 
$M_{200}$ $\geq$ 10$^{14}$ M$_{\sun}$ ram pressure stripping is the dominant process responsible for the quenching of the star formation 
activity of disc galaxies \citep{Boselli+06}. The most convincing evidence
comes from the recent very deep observations of nearby clusters using narrow-band filters centred on the H$\alpha$ 
Balmer recombination line which revealed the presence of extended (up to $\simeq$ 100 kpc) low surface brightness ($\Sigma(\mathrm{H}\alpha)$ $\simeq$ 10$^{-18}$
erg s$^{-1}$ cm$^{-1}$ arcsec$^{-2}$) tails of ionised gas. The cometary shape of the tails and the lack of any associated diffuse tidal structure rule out any
gravitational perturbation which would affect at the same time the gaseous and the stellar component. This undoubtedly states that the ionised gas tail formed after the
interaction of the galaxy interstellar medium (ISM) with the ICM. The most striking examples have been found first in the nearby cluster A1367 by \citet{Gavazzi+01} 
\citep[see also][]{Boselli+14, Gavazzi+17, Consolandi+17} and are now becoming common in other clusters such as Norma \citep{Zhang+13},
Coma \citep{Yagi+07, Yagi+10, Yagi+17, Fossati+12, Gavazzi+18} and Virgo \citep{Yoshida+02, Kenney+08, Boselli+16a, Boselli+18a, Boselli+18b, Fossati+18}. Tails of stripped material have been observed also in \ion{H}{i} \citep{Chung+07, Scott+12}, 
in CO \citep{Jachym+13, Jachym+14, Jachym+17}, and in X-rays \citep{Sun+06, Sun+07, Sun+10}, indicating that the other gas phases can also be 
perturbed during the interaction. Ram pressure stripping as the dominant mechanism is also suggested by the relative distribution in clusters of gas deficient 
and star forming galaxies \citep{Gavazzi+13, Boselli+14a}, by the outside-in radial truncation of the 
different components of the ISM \citep{Cayatte+90, Cayatte+94, Cortese+10, Cortese+12, Boselli+14b} and of the star forming disc
\citep{Koopmann+01, Boselli+06, Boselli+06a, Boselli+15, Fossati+13}, and by the abrupt
truncation of the star formation activity of cluster galaxies \citep{Boselli+16b, Fossati+18}.

The statistical importance of ram pressure stripping, which is proportional to $\rho_\mathrm{ICM}V^2$ 
(where $V$ is the velocity of the galaxy within the cluster and $\rho_\mathrm{ICM}$ is the density of the ICM), 
is expected to reduce at earlier epochs, when clusters of galaxies 
were first assembling through the accretion of smaller groups \citep{Gnedin03, DeLucia+12}.
Here the reduced density of the ICM and lower velocity dispersion typical of lower mass dark matter halos would rather have favoured gravitational 
interactions able to modify the structural properties of the perturbed galaxies. These gravitationally perturbed objects are probably
the progenitors of massive lenticulars observed in nearby clusters \citep[e.g.][]{Boselli+06}. 
The transformation of galaxies at higher redshift via gravitational perturbations in lower density structures before their accretion 
into massive clusters (pre-processing) is also corroborated by the typical age of the stellar populations of lenticulars, by their perturbed star formation history, and
by their distribution in high-density regions at different epochs \citep{Dressler+97, Dressler04, Poggianti+99, Poggianti+06}.

Despite the aforementioned evolutionary picture is now becoming clearer, we still do not know at which epoch and under which conditions  
gravitational perturbations are superseded by hydrodynamic interactions of galaxies with the ICM.
Although massive clusters were significantly less numerous in the past, the physical conditions encountered by infalling systems were similar to those 
present in local clusters, characterised by similar $\rho_\mathrm{ICM}$ gas densities \citep{Giodini+13, McDonald+17}. It is thus conceivable that 
ram pressure episodes were already present at early epochs.
Because of obvious observational limits, direct evidence for ongoing ram pressure stripping at higher redshift is still lacking. 
\citet{Cortese+07} discovered two galaxies in two clusters at $z$ $\simeq$ 0.2 with extended tails of star forming regions, now generally 
referred to as jellyfish galaxies \citep[e.g.][]{Poggianti+17}. Since then, another $\sim$ fifty objects with similar characteristics 
have been found in deep HST images of clusters at 0.3 $\leq$ $z$ $\leq$ 0.7 \citep{Owers+12, Ebeling+14, McPartland+16}. Since selected
in the optical bands, however, where the emission is dominated by stars of intermediate age, a jellyfish morphology is not a direct proof
of an ongoing ram pressure stripping episode \citep{Cortese+07}.
Indeed, in ram pressure stripped galaxies star formation in the tail is not ubiquitous \citep[e.g.][]{Boselli+16a}, and whenever observed, it is 
generally localised in very compact \ion{H}{ii} regions dominated by young stellar populations emitting mainly in the UV bands. Extended and asymmetric low surface 
brightness structures in the optical bands are due to tidal tails. The only clear example of galaxies undergoing ram pressure stripping 
at intermediate redshift comes from deep Subaru narrow-band imaging observations of the cluster Abell 851 at $z$=0.4 \citep{Yagi+15}.
Indirect evidences at $z$=0.7 comes from the reduced gas content \citep{Betti+19} and
from the extended H$\alpha$ ionised gas emission \citep{Vulcani+16} observed in galaxies located in high-density regions.

The search for galaxies undergoing a ram pressure stripping event in high-$z$ clusters is now possible thanks to the advent of new generation
instruments in the radio millimetric and in the optical domain. The extraordinary sensitivity of ALMA allows the detection of low column 
density tails of molecular gas stripped from perturbed cluster galaxies. The results obtained in the local Universe, however, suggest
that the cold gas, once stripped and in contact with the hot surrounding ICM, rapidly changes phase becoming ionised via different possible mechanisms 
such as heat conduction, shocks, or magneto hydrodynamic waves
\citep{Tonnesen+10, Tonnesen+12, Boselli+16a, Gavazzi+18}. The ionised gas phase can be detected by IFU spectrographs 
provided that their sensitivity is sufficient to reach surface brightnesses as low as $\sim$ 10$^{-18}$ erg s$^{-1}$ cm$^{-1}$ arcsec$^{-2}$,
the typical limiting surface brightness of MUSE after $\sim$ 1 h of exposure.

As part of the MUSE Guaranteed Time Observations (GTO), we have recently undertaken a spectroscopic survey of intermediate 
redshift (0.25 $< z <$ 0.85) groups and clusters (MAGIC: MUSE-gAlaxy Groups In Cosmos, Epinat et al. in prep).
During this survey we detected two star forming galaxies in the COSMOS cluster 32 \citep{Knobel+12} at $z$=$0.73$ with extended (up to $\simeq$ 100 kpc in projected distance) tails of ionised gas
without any stellar counterpart in the deep optical images. This observation is thus the first evidence of an ongoing ram pressure stripping event
at $z$ $>$ 0.5. This paper is structured as follows: in Sect. 2 we briefly describe the cluster, in Sect. 3 the 
observations and the data reduction. The analysis of the MUSE data is given in Sect. 4, and the results are discussed in Sect. 5 and summarised in the conclusions.
We assume a $\Lambda$CDM cosmology with H$_0=70$ km s$^{-1}$, $\Omega_M=0.7$, and $\Omega_\Lambda=0.3$.

\section{The galaxies ID345 and ID473 in the cluster CGr32}

\begin{figure*}
   \centering
   \includegraphics[width=1\textwidth]{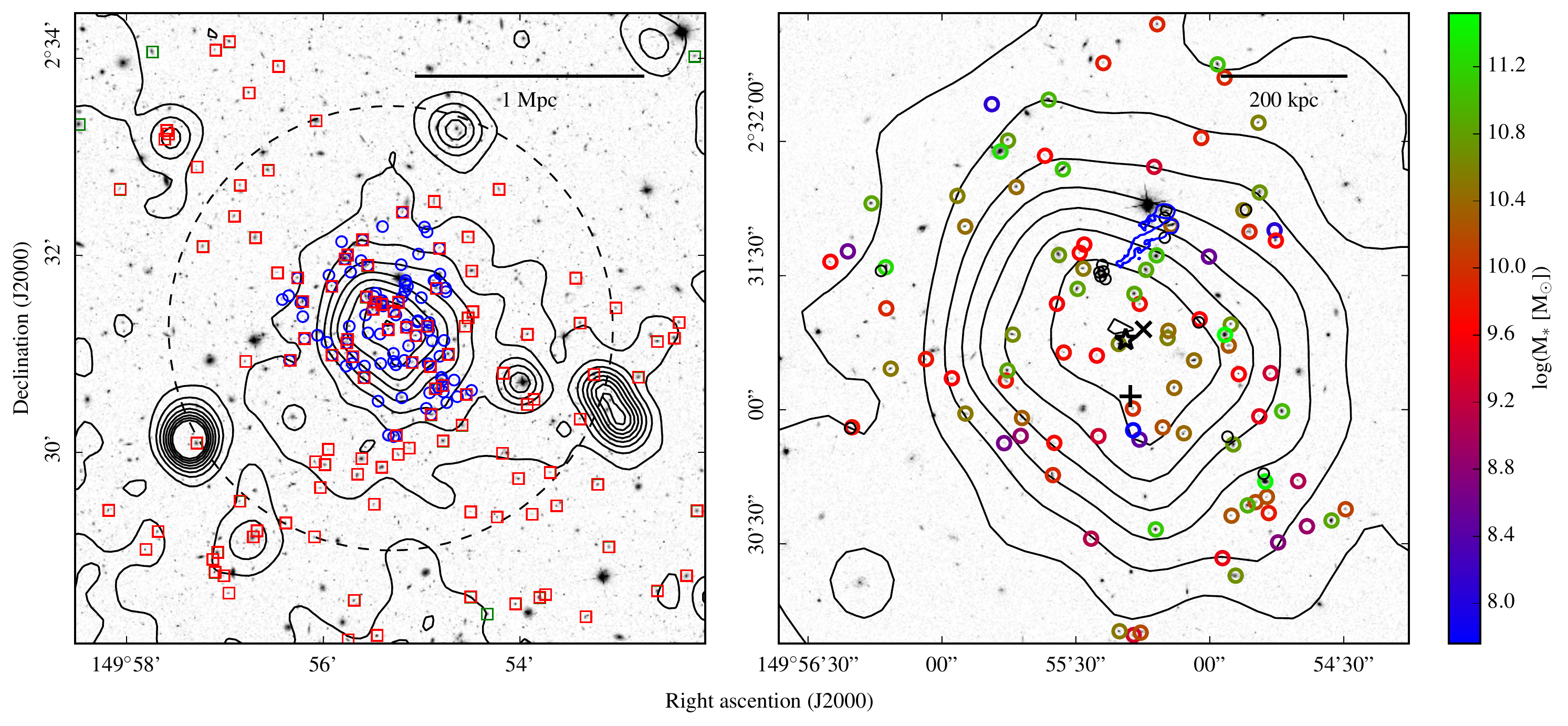}
   \caption{The distribution of galaxies within the cluster CGr32. Left: 6$\arcmin$ $\times$ 6 $\arcmin$ HST/ACS (F814W filter, logarithmic scale, arbitrary units) map of 
    the COSMOS region including the CGr32 cluster of galaxies. Spectroscopically identified cluster members within the MUSE data are indicated with blue 
    circles and galaxies in the \citet{Iovino+16} and \citet{Knobel+12} catalogues within the same redshift range ($0.71685 \le z \le 0.74378$)
    are displayed with red squares. 
    Contours show the X-rays gas distribution \citep[XMM-\textit{Newton}, 0.5-2 keV,][]{Finoguenov+07}, smoothed with a 2 pixel Gaussian. The grey dashed circle represents 
    $R_{200}$.
   Right: zoom on the region mapped with MUSE. Galaxies are colour coded according to their stellar mass (Epinat et al. in prep.) and those 
   without mass estimate due to blending in broad-band images are shown with black circles. The black star indicates the barycentre of  the MUSE cluster members, 
   the black plus sign is the centre quoted by \citet{Iovino+16} and the black cross indicates the X-ray centre of \citet{Gozaliasl+19}.
   The tails of ionised gas are indicated with a blue contour ($\Sigma([\ion{O}{ii}])$ = 1.5 $\times$ 10$^{-18}$ erg s$^{-1}$ cm$^{-2}$ arcsec$^{-2}$).
%
 }
   \label{XY}%
   \end{figure*}

\begin{figure}
   \centering
   \includegraphics[width=0.48\textwidth]{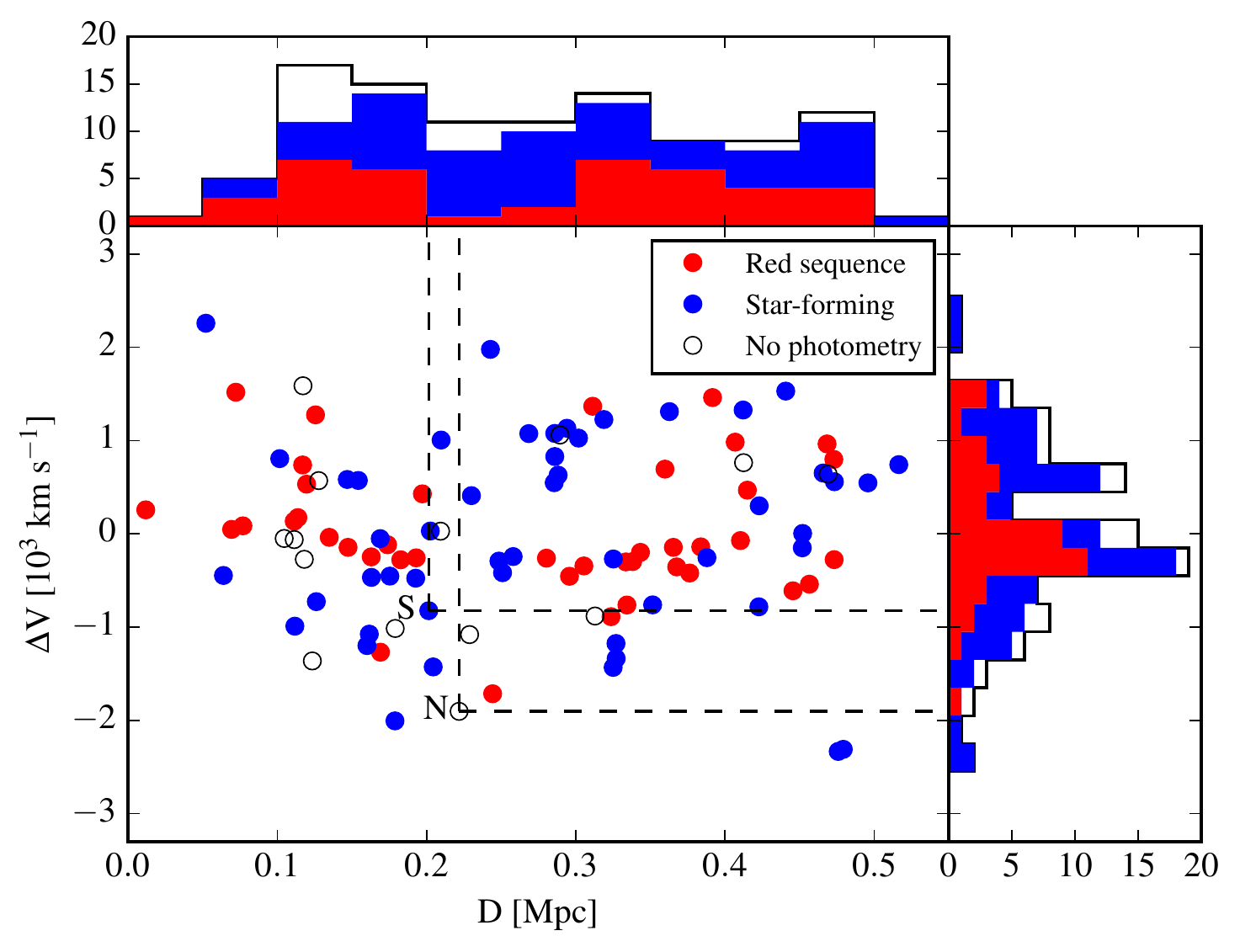}
   \caption{Phase-space diagram of galaxies in the cluster CGr32 determined assuming as centre the barycentre of spectroscopically identified members with MUSE.
   Quiescent galaxies are indicated with red filled dots, star-forming objects with blue filled dots. Black open circles show blended galaxies with 
   unavailable broad-band photometry. The galaxies ID345 and ID473 are indicated with "S" and "N", respectively.
 }
   \label{histo}%
   \end{figure}

Initially identified as a group, the COSMOS cluster 32 \citep[CGr32;][]{Knobel+12} shows properties typical of cluster of galaxies.
CGr32 is located at a redshift of $z$=0.73 and is part of the COSMOS Wall described by \citet{Iovino+16} (see Fig. \ref{XY}). 
Thanks to the new MUSE observations 105 galaxies have been identified as cluster members.
We determined cluster membership based on the spatial and redshift distributions thanks to a friends-of-friends algorithm using a linking length of 450 kpc and a velocity separation of 500~km s$^{-1}$ (Epinat et al., in prep.).
The cluster has a velocity dispersion of $\simeq$ 930 km s$^{-1}$ using all galaxies detected with MUSE, a radius $R_{200}$ $\simeq$ 1.0 Mpc, and a dynamical mass 
$M_{200}$ $\simeq$ 2.3 $\times$ 10$^{14}$ M$_{\sun}$ inferred from X-rays observations \citep[ID 10220 in][]{Gozaliasl+19} and is thus comparable to a fairly 
massive cluster such as Virgo in the local Universe.
The fraction of star forming galaxies ($\sim$ 50 \%)\footnote{The distinction between quiescent and star-forming galaxies is performed from their distribution in the main 
sequence diagram as suggested by \citet{Barro+17},
and consistently with \citet{Boselli+14a}. Stellar masses and star formation rates have been derived as described in Sect. (3.2).} is close to the typical fraction observed 
in similar clusters at that redshift \citep{Dressler+97} and to that of the spiral-rich Virgo cluster \citep[48\%,][]{Boselli+14a}.
The galaxies studied in this work (ID345 and ID473) are two edge-on spirals located at $\simeq$ 200-220 kpc from the cluster centre (see Table \ref{galaxies}) coherently 
identified as the peak of galaxy and of hot gas density in both the MUSE data (Epinat et al., in prep.) and in the XMM-\textit{Newton}/\textit{Chandra} X-rays data 
\citep[][and this work -- see Table \ref{cluster} and Sec~\ref{s:odat}]{Finoguenov+07, George+11, Gozaliasl+19}. The asymmetric velocity distribution 
of galaxy members and a significantly larger velocity dispersion of the star forming galaxies ($\sigma_\mathrm{SF}$ = 1085 km s$^{-1}$) 
with respect to the quiescent objects ($\sigma_\mathrm{Q}$ = 697 km s$^{-1}$) suggest infall into the cluster \citep{Colless+96},
as indeed expected for a cluster still in formation (Fig. \ref{histo}). The galaxies ID345 and ID473 (Fig. \ref{tail}) have a relative line-of-sight velocity with respect to the cluster 
of $\Delta V_\mathrm{ID345} = -824$ km s$^{-1}$ and $\Delta V_\mathrm{ID473} = -1903$ km s$^{-1}$, and are thus crossing the cluster from the back.
Despite its large line-of-sight velocity, ID473 has a high probability to be a real member of the cluster. Indeed, using more stringent parameters for the friends-of-friends algorithm as low as 375 kpc and 400 km~s$^{-1}$, as suggested by \citet{Iovino+16}, it still remains a member.

\begin{table}
\caption{Properties of the cluster CGr32}
\label{cluster}
\begin{tabular}{cccc}
\hline
\noalign{\smallskip}
\hline
Variable        & Value                 & Ref.	& Comment          \\      
\hline
R.A. (J2000)	& 149\degr55\arcmin19.0\arcsec & TW	& MUSE barycentre\\
Dec. (J2000)	& 2\degr31\arcmin15.5\arcsec   & TW	& MUSE barycentre\\
R.A. (J2000)	& 149\degr55\arcmin14.8\arcsec & 1	& from X-rays data\\
Dec. (J2000)	& 2\degr31\arcmin18.0\arcsec   & 1	& from X-rays data\\
$z$		& 0.7303			& 1	&	\\
$\sigma$	& 931 km s$^{-1}$	& TW	& full sample	\\
$\sigma_\mathrm{Q}$	& 697 km s$^{-1}$	& TW	& quiescent galaxies	\\
$\sigma_\mathrm{SF}$	& 1085 km s$^{-1}$	& TW	& star forming galaxies	\\
$R_{200}$	& 1.53 Mpc		& TW	& from MUSE data	\\
$M_{200}$	& 8.14 $\times$ 10$^{14}$ M$_{\sun}$& TW	& from MUSE data$^a$\\
$R_{200}$	& 0.983 Mpc		& 1	& from X-rays data\\  
$M_{200}$	& 2.33 $\times$ 10$^{14}$ M$_{\sun}$& 1	& from X-rays data\\
\noalign{\smallskip}
\hline
\end{tabular}
\\
References: TW: this work, 1) \citet{Gozaliasl+19}.\\
$^a$: This mass reduces to $M_{200} = 3.41 \times 10^{14}$~M$_{\sun}$ if derived using the velocity dispersion of the quiescent population only rather than the full sample.
\end{table}

\begin{table*}
\caption{Properties of the two galaxies with tails}
\label{galaxies}
\begin{tabular}{ccc}
\hline
\noalign{\smallskip}
\hline
Variable        	& ID345                 					& ID473	 	\\
\hline
R.A. (J2000)		& 149\degr55\arcmin8.6\arcsec 						& 149\degr55\arcmin9.7\arcsec		\\
Dec. (J2000)		& 2\degr31\arcmin41.2\arcsec						& 2\degr31\arcmin44.6\arcsec		\\
$V_\mathrm{gal}-V_\mathrm{CGr32}$	& -824 km s$^{-1}$						& -1903	km s$^{-1}$	\\
$R^a$			& 201 kpc							& 222 kpc	\\
$M_\mathrm{star}$		& 2.0 $\times$ 10$^{10}$	M$_{\sun}$			& $\sim$ 10$^{10}$	M$_{\sun}$	\\
$M_\mathrm{dyn}^b$		& 1.0 $\times$ 10$^{11}$	M$_{\sun}$				& 9.7 $\times$ 10$^{10}$	M$_{\sun}$	\\
$\sigma$		& $131 \pm 48$ km s$^{-1}$					& $41 \pm 34$ km s$^{-1}$	\\
SFR			& 38 M$_{\sun}$ yr$^{-1}$ $^c$					& 10-24 M$_{\sun}$ yr$^{-1}$ $^d$\\
tail proj. length	&  97 kpc							& 34 kpc		\\
tail proj. width	&  21 kpc							& 15 kpc		\\
$f([\ion{O}{ii}])_\mathrm{tail}$	& 113.4$\pm$2.3 $\times$ 10$^{-18}$ erg s$^{-1}$ cm$^{-2}$	& 26.4$\pm$1.2 $\times$ 10$^{-18}$ erg s$^{-1}$ cm$^{-2}$ \\
\noalign{\smallskip}
\hline
\end{tabular}
\\
Notes: $^a$ projected distance from the cluster centre. $^b$ Dynamical masses are estimated using the extent measured in the [\ion{O}{ii}] distribution and the velocity amplitude of a rotating disc model which takes 
into account the beam smearing as done in \citet{Contini+16}. The velocity dispersion is corrected for beam smearing and for the instrumental spectral resolution. 
$^c$ Possibly overestimated due to the presence of an AGN.
$^d$ The first value has been measured from the H$\beta$ line using a dust attenuation estimate derived from the Balmer decrement, the second one has been derived from 
the observed [\ion{O}{ii}] line using the calibration of \citet{Moustakas+06}.
\end{table*}

\section{Observations and data reduction}

\subsection{MUSE spectroscopy}

Details of the MUSE observations and data reduction will be presented in a dedicated paper (MAGIC survey paper, Epinat et al. in prep). Here we briefly summarise the main steps.
The observations of CGr32 were obtained during MUSE-GTO as part of the MAGIC project (Epinat et al. in prep, PI: T. Contini). Three fields were necessary to map 
the cluster core. An equal observing time of 4.35h was spent on each field over three observing runs: 1 hour per field without adaptive optics (AO) in April 2016 
(Program ID 097.A-0254), 3.3 hours per field with AO in March (3.3 hours in total, Program ID 100.A-0607) and April 2018 (6.7 hours in total, Program ID 101.A-0282). 
For each run, the observing block sequence consisted in four 900 seconds exposures each taken after rotating the field by 90 degrees.

For each of the three fields, we produced a single reduction combining all exposures with and without AO. The reduction was performed using the 
MUSE standard pipeline v1.6 \citep{Weilbacher+12, Weilbacher+14, Weilbacher15}.
The data reduction process generates data cubes with a spatial sampling of 0.2\arcsec\ and a spectral sampling of 1.25 \AA. The spatial extent of each field is of
one square arcminute and the spectrum ranges from 4750 \AA ~to 9350 \AA.

The point spread function (PSF) of the combined exposures was estimated on each field from a Gaussian fit to the stars present in each datacube. 
The average full width at half maximum (FWHM) at $\lambda = 6450$ \AA ~ ranges from 0.60\arcsec\ to 0.72\arcsec\ depending on the field. The two galaxies with 
tails are visible on the same field which has a FWHM of 0.60\arcsec.

We derived [\ion{O}{ii}]\footnote{Hereafter [\ion{O}{ii}] refers to the [\ion{O}{ii}]$\lambda\lambda$3727-3729\AA~ emission-line doublet.} flux (Fig. \ref{tail}) and kinematics (Fig. \ref{kin}) maps for the two 
galaxies with tails using \textsc{Camel}\footnote{\url{https://gitlab.lam.fr/bepinat/CAMEL.git}} \citep{Epinat+12}, 
which fits any emission line using a Gaussian function and a polynomial continuum to any pixel 
of a datacube, after a Gaussian spatial smoothing with a FWHM of $3\times 3$ pixels and using a continuum of degree one, mainly to account for 
the continuum of the strong star in the field which is close to the two galaxies.
We also extracted the same maps with a Gaussian spatial smoothing of $4\times 4$ pixels FWHM in order to
define masks where the tails are detected in [\ion{O}{ii}]. 
The masks were defined using both a flux threshold of $7.5\times 10^{-19}$~erg~s$^{-1}$~cm$^{-2}$~arcsec$^{-2}$ (1.5 times the surface brightness limit in the smoothed [\ion{O}{ii}] map) and a criterion on kinematics of the detected structure to remove regions with velocity discontinuities larger than $\approx 200$~km~s$^{-1}$. 
An additional mask was applied to hide the emission from other galaxies in the group. These final masks were used to produce integrated spectra (and their variance 
spectra) for both galaxies and associated tails (cf. Fig. \ref{spectra}).
The velocity amplitude over these galaxies and tails is $\sim 400$~km~s$^{-1}$, which would broaden emission lines in the spectra integrated over the masks and thus decrease the signal to noise ratio. Therefore, in order to remove large scale velocity variations, each pixel is set, before integration, at rest wavelengths using the spatially resolved velocity field shown in Fig. \ref{kin}. The residual broadening due to the uncertainties in the velocity field is $\sim 30-50$~km~s$^{-1}$, which is below the spectral resolution of the MUSE data \citep[see ][]{Bacon+17} over the whole spectral range shown in Fig. \ref{spectra}.
The continuum due to the bright star was removed from the galaxy ID473 and its
tail spectra by scaling the absorption lines of the star spectrum. For the tail a residual continuum signal due to both the star and a background galaxy was also removed.
In the tails the [\ion{O}{ii}] doublet flux and 
its uncertainty was then measured as the sum of the spectra around a narrow spectral window centred on the line which corresponds to a velocity amplitude of 
600 km~s$^{-1}$ around each side of the lines. 
Uncertainties were determined over the same range from the variance spectrum.
The observed fluxes in the tail are given in Table \ref{galaxies}. 
We also measured the mean [\ion{O}{ii}] doublet ratio within the tail. The mean ratio in the tail of ID345, the only one
where the signal is sufficiently high to make this velocity correction possible, is 
[\ion{O}{ii}]$\lambda$3729/[\ion{O}{ii}]$\lambda$3726 = 1.15 $\pm$ 0.06. We recall, however, that this is a light-weighted mean, probably biased
towards the emission of the brightest regions, possibly associated to compact but unresolved \ion{H}{ii} regions. If we try to measure the [\ion{O}{ii}] doublet 
ratio pixel per pixel along the tail, its ratio is [\ion{O}{ii}]$\lambda$3729/[\ion{O}{ii}]$\lambda$3726 ~ $\simeq$ 1.5 in more than 50\% of the pixels.
We also tried to measure H$\beta$ and [\ion{O}{iii}]$\lambda$5007\AA ~ line fluxes in the tails using the masks defined for the [\ion{O}{ii}] doublet.
We have a $\sim$ 1.5-$\sigma$ detection of H$\beta$ on the integrated spectrum of the tail associated to ID345 
($f_{\mathrm{ID345}}(\mathrm{H}\beta)$ = 7.76$\pm$5.25 $\times$ 10$^{-18}$ ~erg~s$^{-1}$~cm$^{-2}$), and an upper limit in the tail of ID473 
($\sigma_\mathrm{ID473}(\mathrm{H}\beta)$ = 1.92 $\times$ 10$^{-18}$ ~erg~s$^{-1}$~cm$^{-2}$), as well as upper limits in both tails for the [\ion{O}{iii}] line
($\sigma_\mathrm{ID345}([\ion{O}{iii}])$ = 5.23 and $\sigma_\mathrm{ID473}([\ion{O}{iii}])$ = 2.03 in units of 10$^{-18}$~erg~s$^{-1}$~cm$^{-2}$).


For the two galaxies, we first removed the stellar continuum using pPXF \citep{Cappellari+04} with the MILES stellar population synthesis library 
\citep{Falcon-Barroso+11}. We then determined fluxes of H$\beta$, H$\gamma$, [\ion{O}{ii}] and [\ion{O}{iii}] using the same method as for the tails.


\subsection{Other data}
\label{s:odat}

CGr32 is inside the COSMOS field \citep{Scoville+07} and has therefore been observed in many bands. We used the HST-ACS image of the field 
observed with the F814W filter for a morphological analysis of the two galaxies with \textsc{Galfit} \citep{Peng+02}, while the Subprime/Subaru camera image in the $i'$-band \citep{Taniguchi+07},
the deepest available for this field with a limiting surface brightness of 28 mag arcsec$^{-2}$, 
was used to search for any possible red stellar population counterpart in the galaxy tails. We also used the XMM-\textit{Newton} data
for the determination of the density profile of the hot ICM.
For this purpose we used the COSMOS mosaic image in the 0.5 to 2 keV band \citep{Hasinger+07, Finoguenov+07} for illustration 
purposes in Fig.~\ref{XY} and the OBSID 0203361701 archive data for the extraction of the hot gas density distribution. This is the longest exposure 
($\sim$32~ksec) covering CGr32 within the inner 10~arcmin of the FoV. The data were processed following the methodology presented in \citet{Pratt+07} 
and \citet{Bartalucci+17}. The 3D density profile is derived from the surface brightness profile (extracted at the position of the X-ray centre reported 
in Table~\ref{cluster}) using a non-parametric deconvolution and deprojection method with regularisation \citep[][-- see Fig.~\ref{X_profile} and Sec.~\ref{s:dis}]{Croston+06}.

We determined galaxy parameters, such as stellar mass, star formation rate (SFR), and extinction, from spectral energy distribution models using the 
FAST code \citep{Kriek+09} as described in \citet{Epinat+18} for a similar dataset in the COSMOS field with deep imaging data from the UV to the near-IR.
Stellar masses and SFR are calculated assuming a \citet{Chabrier03} IMF. 
The spectrum of ID345 exhibits
asymmetric shapes of the emission lines [\ion{O}{iii}]$\lambda\lambda$4959,5007
and H$\beta$ which could be explained by a contribution from an AGN. The presence of an AGN is also suggested by its strong
observed radio emission at 1.4 GHz
(0.115 mJy - \citealp{Schinnerer+10} - corresponding to a rest frame luminosity $L(\mathrm{1.4~GHz})$ = 4 $\times$ 10$^{23}$ W~Hz$^{-1}$ derived assuming a nonthermal spectral index $\alpha = 0.8$ - \citealp{Condon92}).
Should such an AGN indeed be present, this could bias the SFR derived from SED fitting. For the galaxy ID473, where 
the photometry is highly contaminated from a nearby star, the stellar mass is just a rough estimate. For this galaxy the SFR is determined using two different methods, 
the first one from the H$\beta$ emission line corrected for dust attenuation using
the Balmer decrement, the second one from the observed [\ion{O}{ii}] line using the calibration of \citet{Moustakas+06}.

\section{Analysis}

\subsection{Physical properties}

\begin{figure}
   \centering
   \includegraphics[width=0.48\textwidth]{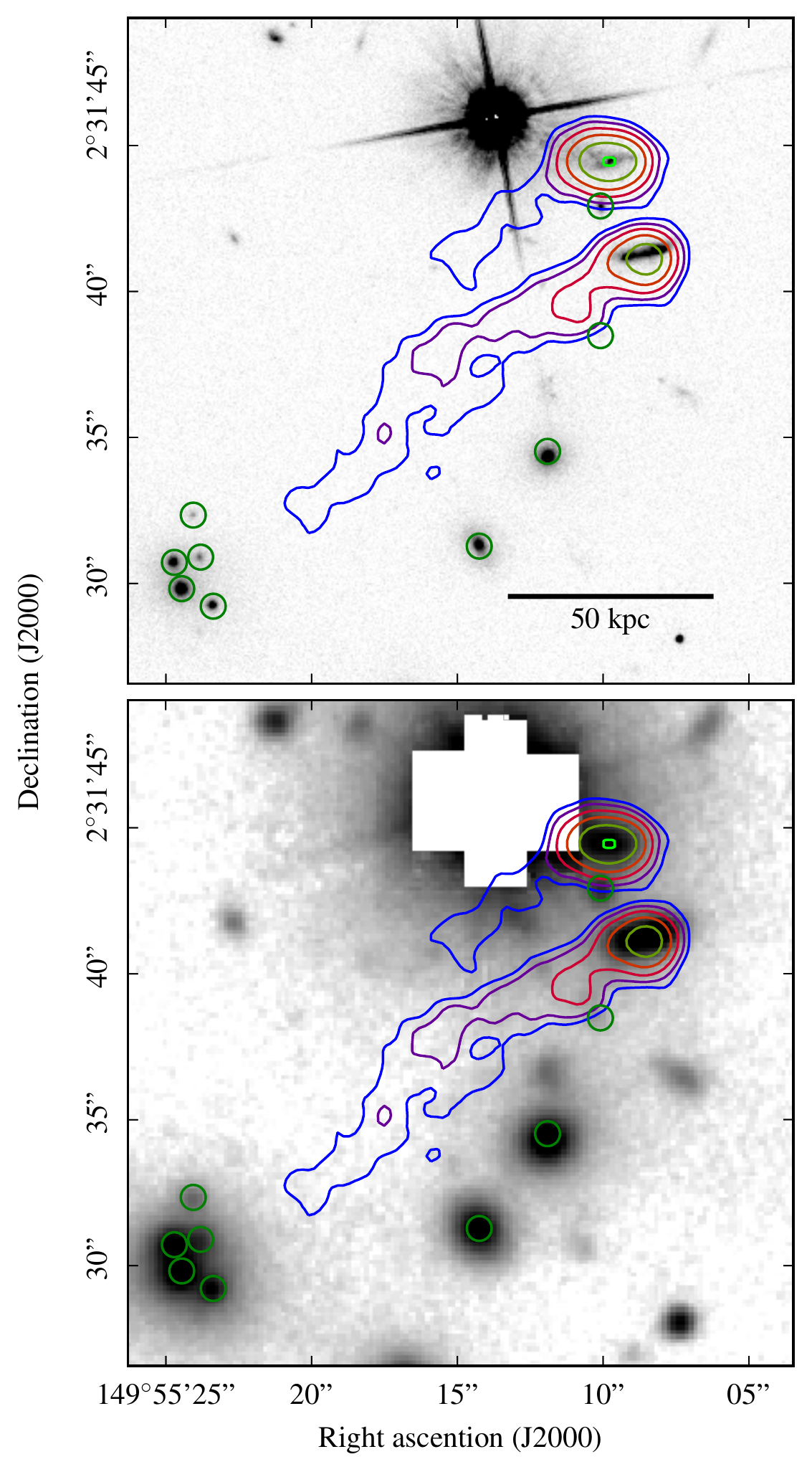}
   \caption{The F814W HST/ACS (top) and the $i'$-band Subaru (bottom) images of the galaxies ID345 (south) and ID473 (north) in the cluster CGr32. Other cluster members are shown with green circles. Contours show the [\ion{O}{ii}] 
   flux map obtained using a Gaussian smoothing of 3 pixels FWHM at a level of $\Sigma([\ion{O}{ii}])$ = 1.5, 3.3, 7.4, 16.3, 36.1, 80.0 $\times$ 10$^{-18}$ erg s$^{-1}$ cm$^{-2}$ arcsec$^{-2}$.
   Both images (oriented as in the sky, north is up, east on the left) show the lack of any optical counterpart along the tail.
 }
   \label{tail}%
   \end{figure}

\begin{figure*}
   \centering
   \includegraphics[width=0.96\textwidth]{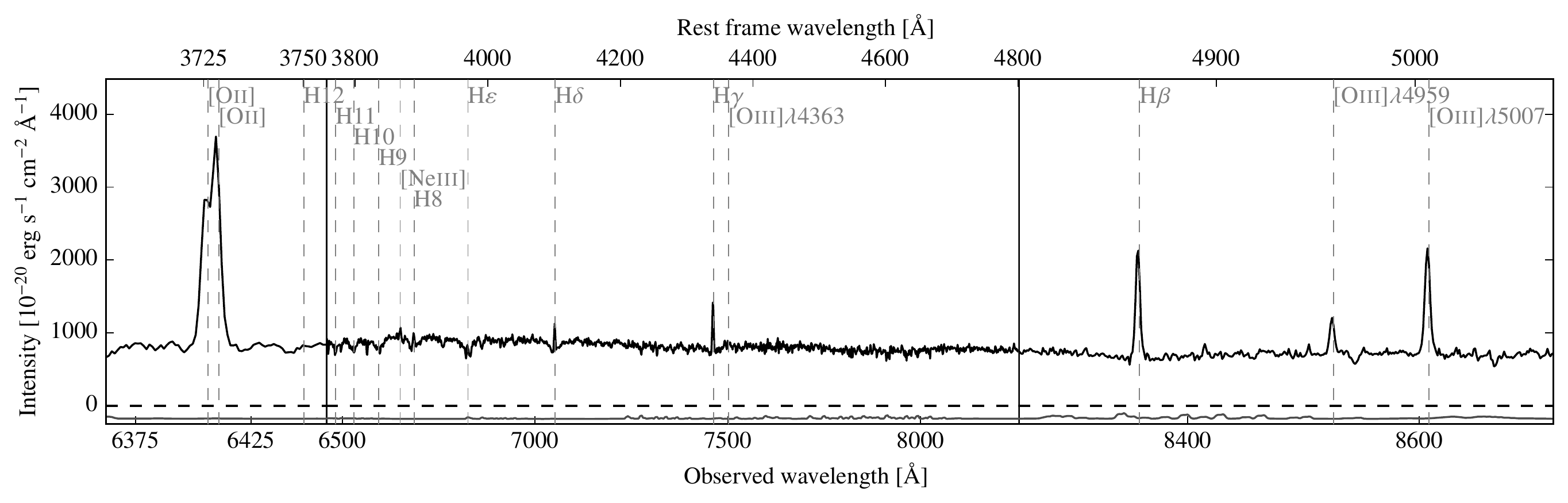}\\
   \includegraphics[width=0.96\textwidth]{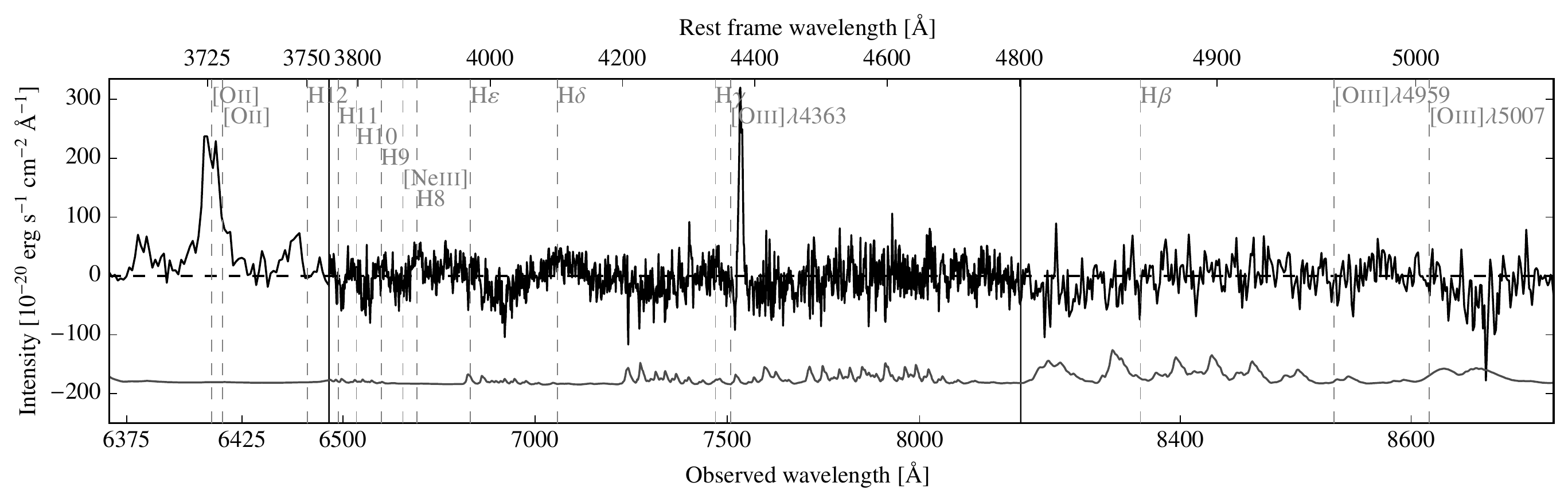}\\
   \includegraphics[width=0.96\textwidth]{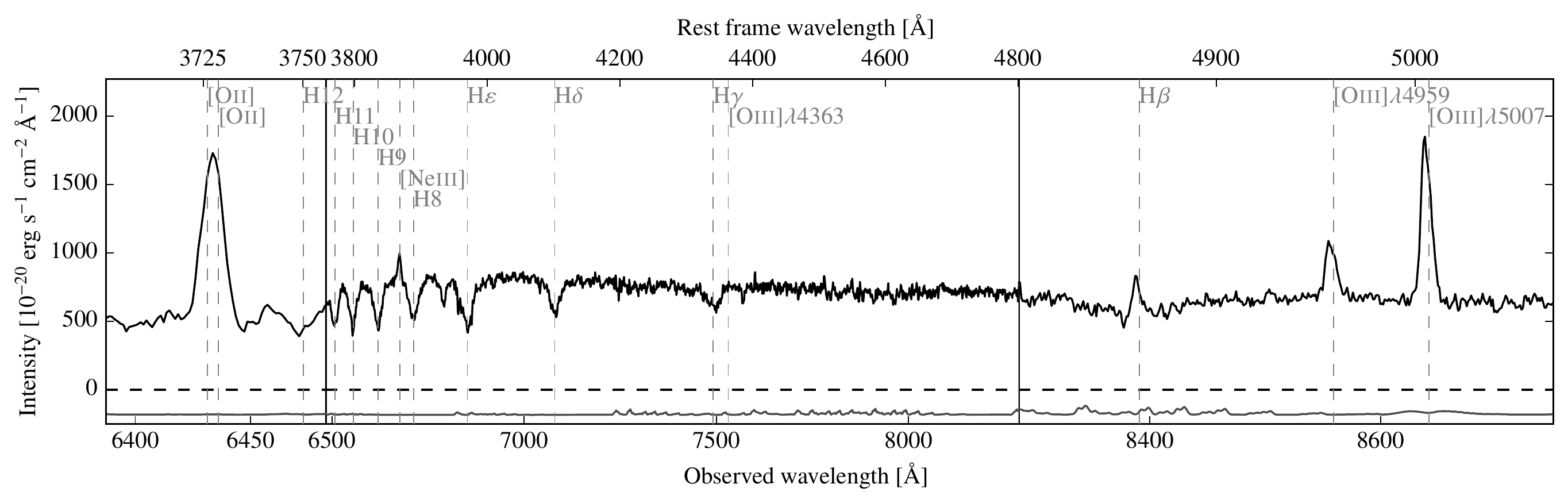}\\
   \includegraphics[width=0.96\textwidth]{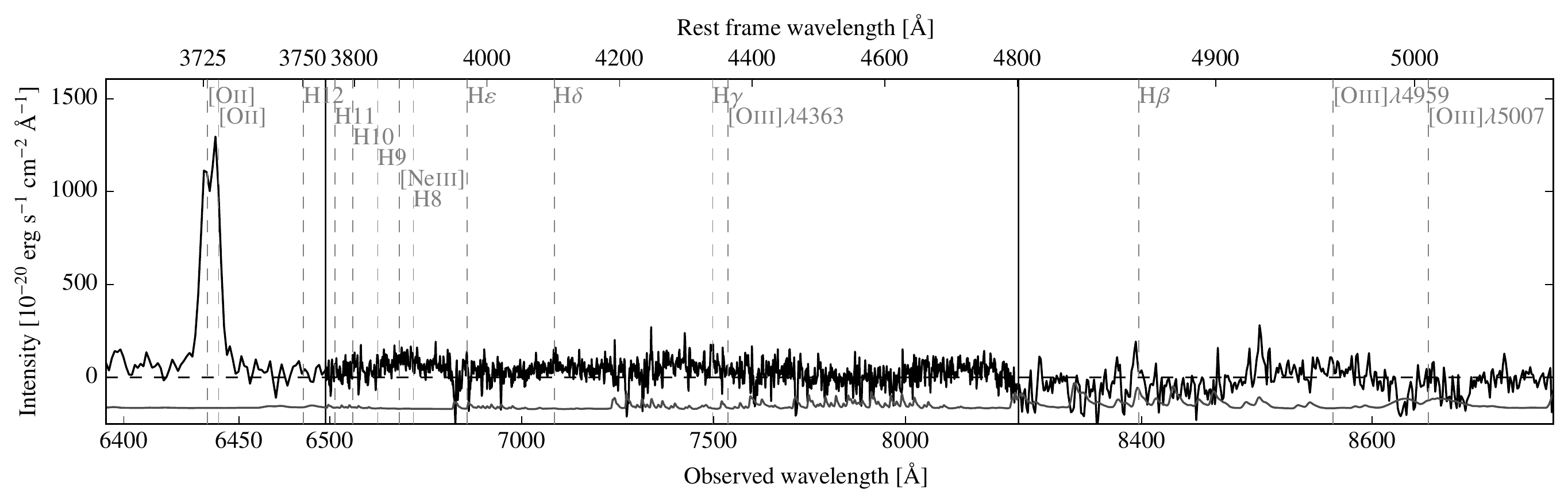}\\
   \caption{The MUSE integrated spectra corrected for variations in the velocity field of galaxy ID473 (upper panel), of its tail (upper middle panel), of galaxy ID345 (lower middle panel), and of its tail (lower panel). 
   The associated standard deviation, which is shifted below zero for clarity, is shown in grey. The main lines at the redshift of the sources are indicated with grey 
   dotted vertical lines. An emission around 7500 \AA\ associated to a background [\ion{O}{ii}] emitter is observed in the spectra of ID473's tail. Fluctuations in the spectra of tails are due to imperfect bright star continuum subtraction and to faint background sources.
 }
   \label{spectra}%
   \end{figure*}

The MUSE observations of the cluster reveal the presence of an extended tail of ionised gas detected in the [\ion{O}{ii}] line associated to the two galaxies ID345 and ID473,
as depicted in Fig. \ref{tail}. The two tails, which extend from the galaxy discs in the south-east direction, are 13.4$\arcsec$ (97 kpc) and 4.7$\arcsec$ (34 kpc) long (projected distance)
and 2.9$\arcsec$ (21 kpc) and 2.1$\arcsec$ (15 kpc) wide (see Table \ref{galaxies}). The deep optical images of the cluster do not show any 
stellar counterpart associated to the extended ionised gas tail (Fig. \ref{tail}) down to a surface brightness limit of $\simeq$ 28 mag arcsec$^{-2}$ 
in the $i'$-band.
The orientation of the tails, which in both cases is parallel to the stellar disc major axis, also indicates that the gas stripping is occurring 
edge-on.
Figure \ref{spectra} shows the spectra of the galaxies and of their tails.

We can compare the properties of the tails to those observed in local galaxies. Unfortunately, at this redshift, the H$\alpha$ line 
is outside the spectral domain of MUSE while H$\beta$ is only barely detected (1.5-$\sigma$ detection) in the tail of ID345 and undetected in ID473, we thus have to derive the intensity of 
the [\ion{O}{ii}] line and convert it into H$\alpha$ assuming a standard [\ion{O}{ii}]/H$\alpha$ ratio.
Spectroscopic observations of the tails of ionised gas in ram pressure stripped galaxies of the local Universe are limited to the spectral domain $\lambda$ $\geq$ 4000-9000 \AA~
\citep{Yoshida+04, Yoshida+12, Fossati+16, Poggianti+17}, preventing any estimate of this ratio from local measurement. The only direct measurement 
of the [\ion{O}{ii}]/H$\alpha$ ratio in the stripped material of local galaxies is the value derived by \citet{Cortese+06} around 
some gravitationally perturbed galaxies in the cluster A1367 ([\ion{O}{ii}]/H$\alpha$ $\simeq$ 2). 
We do not expect that the physical properties of the stripped gas, which is not associated to any stellar component
and is probably embedded in a similarly hot and dense ICM, differ significantly from those observed in A1367. 
This value is consistent with the value derived using the 1.5-$\sigma$ detection of H$\beta$ in the tail of ID345
([\ion{O}{ii}]/H$\beta$ $=$ 14.6) and the 3-$\sigma$ upper limit in ID473 ([\ion{O}{ii}]/H$\beta$ $\geq$ 4.6)
and adopting H$\alpha$/H$\beta$ = 2.86 if we make the reasonable assumption that the dust attenuation is negligible. 
We thus derive the H$\alpha$ luminosity assuming [\ion{O}{ii}]/H$\alpha$ = 2
($L(\mathrm{H}\alpha)$ = 1.3 $\times$ 10$^{41}$ erg s$^{-1}$ for ID345 and  $L(\mathrm{H}\alpha)$ = 2.5 $\times$ 10$^{40}$ erg s$^{-1}$ for ID473) and 
the mean surface brightness of the tails ($\Sigma(\mathrm{H}\alpha)$ = 1.5 $\times$ 10$^{-18}$ erg s$^{-1}$ arcsec$^{-2}$ for ID345 and 
$\Sigma(\mathrm{H}\alpha)$ = 1.3 $\times$ 10$^{-18}$ erg s$^{-1}$ arcsec$^{-2}$ for ID473). 
These numbers, which are close to the detection limits of modern instrumentation, are comparable to those observed in the nearby Universe. 
Due to cosmological surface brightness dimming, however, when transformed to physical units both $H\alpha$ luminosities and surface brightnesses are a factor of $\sim$ 10 higher than those
measured in local systems.

We can also use the [\ion{O}{ii}]$\lambda$3729/[\ion{O}{ii}]$\lambda$3726 line ratio to derive the gas density within the tails. For a mean value of
[\ion{O}{ii}]$\lambda$3729/[\ion{O}{ii}]$\lambda$3726 = 1.15 $\pm$ 0.06 measured on the integrated spectrum of the tail of ID345, the corresponding electron density is $n_e$ $\simeq$ 200 cm$^{-3}$.
As mentioned in Sect. 3.1, however, this is a light-weighted mean, and is thus probably biased towards the brightest and highest density regions, unresolved in the MUSE data.
Indeed, this gas density perfectly matches that observed in star forming regions of galaxies at $z$ $\simeq$ 2 \citep{Sanders+16}. The typical line ratio measured pixel per pixel in the diffuse tail
is rather [\ion{O}{ii}]$\lambda$3729/[\ion{O}{ii}]$\lambda$3726 $\simeq$ 1.5, corresponding to a gas density $n_e$ $\lesssim$ 10 cm$^{-3}$ \citep{Osterbrock+06}.
This upper limit is consistent with the gas density derived from the [\ion{S}{ii}]$\lambda$6716/[\ion{S}{ii}]$\lambda$6731 line ratio measured in the diffuse tails of local galaxies,
or independently with geometrical considerations \citep[e.g.][]{Fossati+16, Boselli+16a}.
%
As in \citet{Boselli+16a} and \citet{Fossati+16}, we can also estimate the typical recombination time of the ionised gas,

\begin{equation}
\tau_\mathrm{rec} = \frac{1}{n_e \alpha_A}
\end{equation}

\noindent
where $\alpha_A$ is the total recombination coefficient ($\alpha_A$ = 4.2 $\times$ 10$^{-13}$ cm$^3$ s$^{-1}$; \citealp{Osterbrock+06}).
For $n_e$ = 10 cm$^{-3}$, $\tau_\mathrm{rec}$ $\simeq$ 10$^4$ yr, a very short time compared to the travel time of the tail 
($\tau_\mathrm{travel}$ $\simeq$ 10$^8$ yr).

This very short recombination time suggests that, as in local galaxies, a source of gas excitation within the tail must be present.
In the lack of photoionisation, which requires the presence of star forming regions within the tail, other ionising mechanisms can be 
thermal conduction, magneto hydrodynamics waves, and shocks \citep{Fossati+16, Boselli+16a}.
The presence of shock-ionised gas is suggested by the high [\ion{O}{ii}]/[\ion{O}{iii}]$\lambda$5007 ratio observed in the tails 
([\ion{O}{ii}]/[\ion{O}{iii}]$\lambda$5007 $\gtrsim$ 7 and
4.5 in ID345 and ID473, respectively; \citealp{Dopita+95}).
Part of the stripped gas might still be in other phases such as cold \ion{H}{i} or hot gas emitting in X-rays, as indeed suggested by simulations \citep{Tonnesen+10, Tonnesen+12}.


   \begin{figure}
   \centering
   \includegraphics[width=0.48\textwidth]{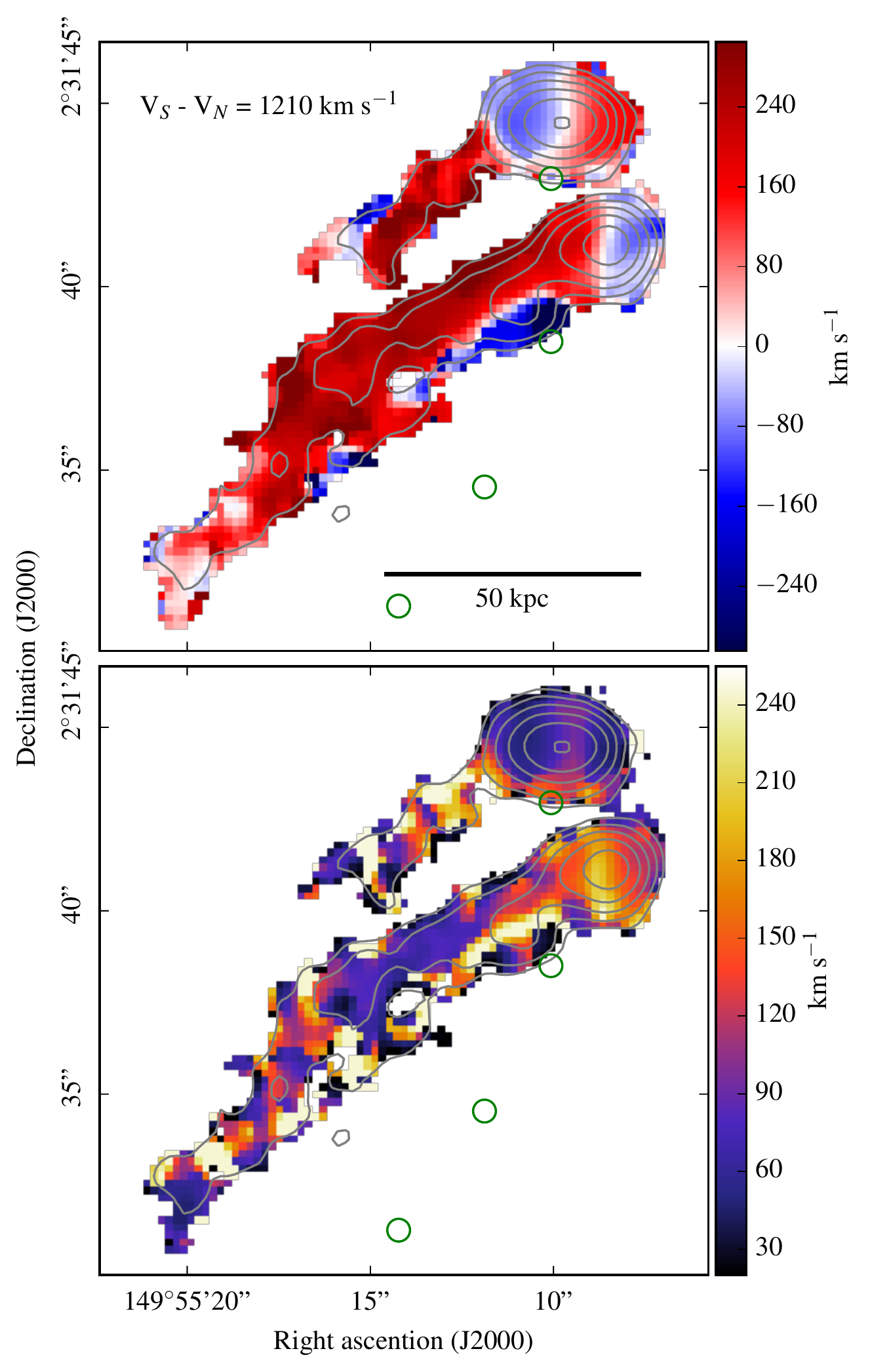}
   \caption{Top: Velocity field of the tails obtained using a Gaussian smoothing of 3 pixels FWHM. The two subsystems have been put at rest frame 
   so that they can be compared on the same figure. Bottom: Velocity dispersion map of the tail obtained using a Gaussian smoothing of 3 pixels FWHM. 
   The contours correspond to [\ion{O}{ii}] flux 
   emission as in Fig. \ref{tail}. The green circles indicate the position of other cluster member galaxies.
  }
   \label{kin}%
   \end{figure}

\subsection{Kinematical properties}

Figure \ref{kin} shows the velocity field and the velocity dispersion map of the ionised gas over the two perturbed galaxies and along the tails.
The velocity field of both galaxies is very symmetric and the difference between the receding and approaching sides of both almost
edge-on rotating ($> 70$\degr) discs is $\sim$ 300 km s$^{-1}$. In addition, the size of the two galaxies is quite similar, which suggests that they have 
a comparable dynamical mass (cf. Table \ref{galaxies}).
The velocity of the gas is fairly constant along the tails and is a bit higher than that of the associated galaxies, which means that the gas is 
decelerated in the tail, confirming its stripped gas origin. This difference is higher for the northern galaxy ID473 ($\Delta V \sim 100$ km s$^{-1}$). 
The low velocities (in blue) associated to a higher velocity dispersion in the southern tail are due to the overlap of the tail with a galaxy of the 
cluster with a lower velocity than the tail (cf. Fig. \ref{tail}).
The velocity dispersion of the gas is high in the southern galaxy ID345 ($\sim$ 131 km s$^{-1}$) which seems to host an active galactic nucleus 
(based on broad lines and a high [\ion{O}{iii}]$\lambda$5007/H$\beta$ ratio around 3.5 on the integrated spectrum).
It is lower ($\sim$ 75 km s$^{-1}$) in the northern galaxy ID473,
while the velocity dispersion of the gas along both tails is fairly high ($\sim$ 80 km s$^{-1}$). We recall, however, that this is a light-weighted estimate of the velocity dispersion of the gas within the tails. It is comparable to the median dispersion along tails ($\sim$ 90 km s$^{-1}$) determined on the velocity dispersion map (see Fig. \ref{kin}), but much more uncertain due to the low signal-to-noise.
Similar velocity dispersions are observed within the tail of local galaxies \citep[e.g. ESO 137-001,][]{Fumagalli+14}.

\section{Discussion}
\label{s:dis}

   \begin{figure}
   \centering
   \includegraphics[width=0.48\textwidth]{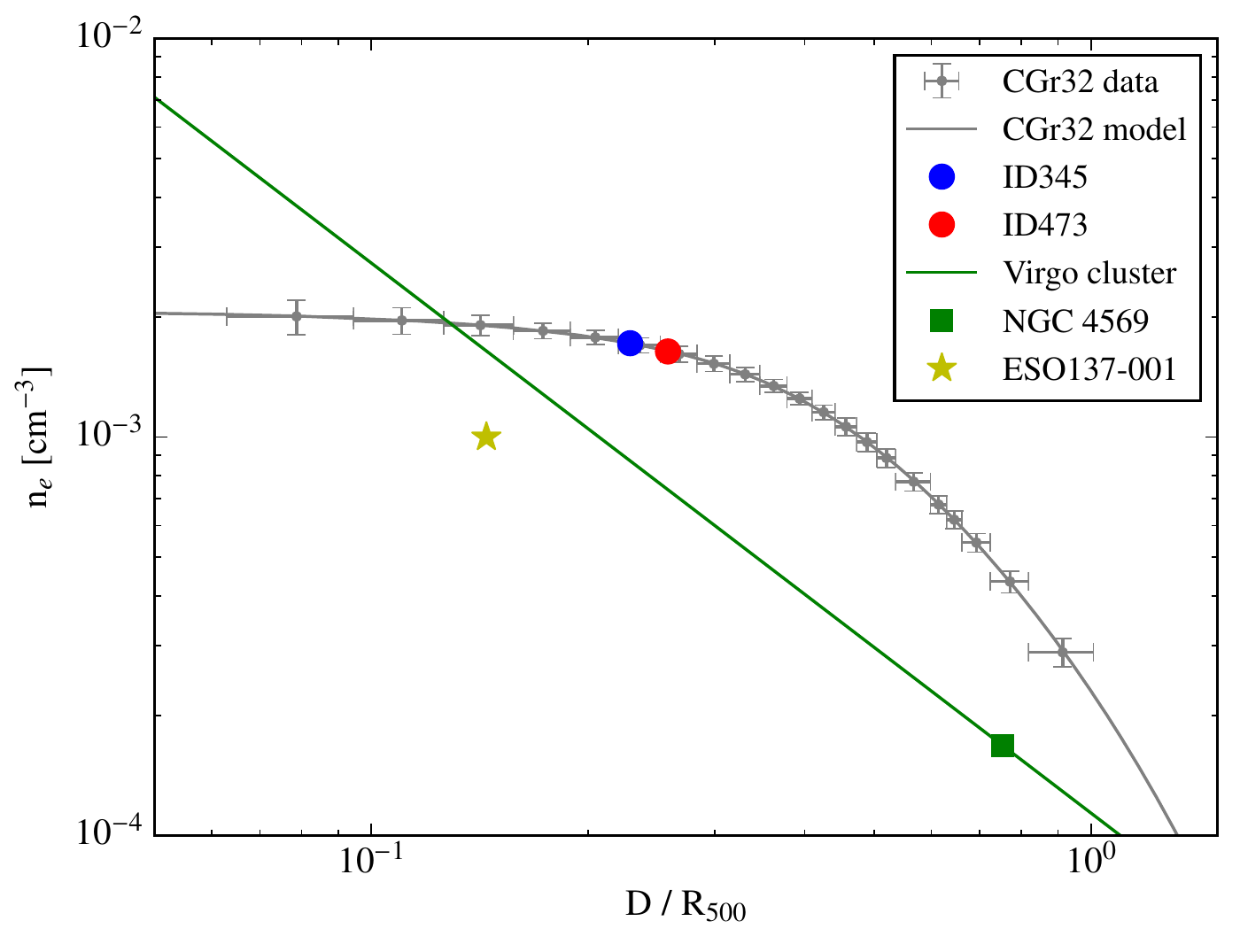}
   \caption{Normalized radial profile of the intra-cluster electron density of CGr32 (data as grey points and best fit as a grey solid line) derived from XMM-\textit{Newton} observation compared to that of 
   the Virgo cluster (green solid line) derived from \textit{Suzaku} and \textit{Planck} data \citep{Simionescu+17}. The blue and red dots, the green square and the yellow star indicate the position of the galaxies ID345, ID473 in CGr32,
   NGC 4569 in Virgo, and ESO 137-001 in Norma, respectively. Norma R$_{500}$ was estimated using \citet{Sun+09} mass-temperature relation with kT~=~6~keV.
   }
   \label{X_profile}%
   \end{figure}

The presence of long tails of ionised gas without any old stellar counterpart suggests that the two galaxies are now undergoing a ram pressure stripping episode.
The derived density, mass, and recombination time of ionised gas stripped during the interaction, as well as the shape and size of the tail 
associated to the two galaxies ID345 and ID473, are very similar to those observed in the massive spiral NGC 4569 in the Virgo
cluster 
or ESO 137-001 in the Norma (A3627) cluster. 
The conditions for gas stripping via ram pressure, $\rho_\mathrm{ICM}V^2$ $>$ 2$\pi$G$\Sigma_\mathrm{gas}$$\Sigma_\mathrm{star}$ \citep{Gunn+72}, 
where $\Sigma_\mathrm{gas}$ and $\Sigma_\mathrm{star}$ are the density of gas and stars over the disc of the galaxy, can be compared to those 
encountered in these nearby counterparts. 
Indeed, the velocity dispersion of the cluster CGr32 ($<V_\mathrm{cluster}>$ $\simeq$ 930 km s$^{-1}$) is close to that of 
Virgo ($<V_\mathrm{cluster}>$ $\simeq$ 800 km s$^{-1}$, \citealp{Boselli+14a}) and Norma ($<V_\mathrm{cluster}>$ $\simeq$ 925 km s$^{-1}$, \citealp{Sun+10}), where ram pressure stripping is active.

%

We use the radial profile of the intracluster electron density, $n_e$, derived from XMM-\textit{Newton} data (see Sec.\ref{s:odat}) to estimate $n_e$ at the position of the two target galaxies (Fig. \ref{X_profile}). 
The azimuthal mean density of the hot intracluster gas at the position of the two galaxies ID345 and ID473 is $n_e$ = 1.7 $\times$ 10$^{-3}$ and $n_e$ = 1.6 $\times$ 10$^{-3}$ cm$^{-3}$, 
respectively. These densities are comparable to those encountered in similar clusters in the redshift range $z$=0.7 to $z$=0 \citep{Giodini+13, McDonald+17}.

These electron densities can also be compared to that measured in the Virgo cluster 
at a similar radial distance from the cluster core ($n_e$ $\simeq$ 2 $\times$ 10$^{-3}$ cm$^{-3}$ at 100 kpc, 
$n_e$ $\simeq$ 10$^{-3}$ cm$^{-3}$ at 200 kpc), or at the distance of NGC 4569 ($n_e$ $\simeq$ 2 $\times$ 10$^{-4}$ cm$^{-3}$ at 500 kpc)\footnote{The electron density of the ICM 
at the distance of NGC 4569 is significantly different than the one reported in \citet{Boselli+16a} which was based on old \textit{ROSAT} data.}
as derived from \textit{Suzaku} and \textit{Planck} data by \citet{Simionescu+17}. It can also be compared to that 
in the Norma cluster at the distance of ESO 137-001 ($n_e$ $\simeq$ 10$^{-3}$ cm$^{-3}$ at 180 kpc, \citealp{Sun+10}).
The conditions encountered by the galaxies ID345 and ID473 in the cluster CGr32, very similar to those observed in similar objects 
in the local Universe, are optimal for an efficient ram pressure stripping event. The gas column density over the disc of the two galaxies is probably higher than those present in ESO 137-001 and NGC 4569 , 
while the stellar density is probably lower just because they are observed at much earlier epochs ($z$ = 0.7). In particular we recall that NGC 4569 
is a bulge dominated Sa galaxy. The relative absolute line-of-sight velocity with respect to the mean velocity of the cluster are $V_\mathrm{ID345}$ $-$ $<V_\mathrm{CGr32}>$ = $-$824 km s$^{-1}$
and $V_\mathrm{ID473}$ $-$ $<V_\mathrm{CGr32}>$ = $-$1903 km s$^{-1}$. 
The projected extension of the ionised gas tails suggests that the galaxies are also moving on the plane of the sky, 
thus these values should be considered as lower limits. These numbers can also be compared to the relative line-of-sight velocity of 
NGC 4569 with respect to Virgo, $V_\mathrm{NGC~4569}$ - $<V_\mathrm{Virgo}>$ = 1176 km s$^{-1}$, ESO 137-001 with 
respect to Norma, $V_\mathrm{ESO~137-001}$ - $<V_\mathrm{Norma}>$ = 191 km s$^{-1}$, and UGC 6697 with respect to A1367, $V_\mathrm{UGC~6697}$ - $<V_\mathrm{A1367}>$ =  132 km s$^{-1}$. 
This simple comparison suggests that, as in their local counterparts, ID345 and ID473
are suffering a ram pressure stripping event. 
It is also worth mentioning that one of the two galaxies, 
ID345, shows prominent Balmer absorption lines (see Fig. \ref{spectra}), typical in objects suffering an abrupt variation of the star formation activity.
Its position above the main sequence (at $z$ = 0.7 main sequence galaxies of $M_{star}$ $\sim$ 10$^{10}$ M$_{\odot}$ 
have star formation rates $SFR$ $\sim$ 10 M$_{\odot}$ yr$^{-1}$, \citealp{Whitaker+14}) suggests that this galaxy might have undergone a moderate starburst phase 
induced by the compression of the gas along the disc \citep[e.g.][]{Bekki14}. The decrease of activity indicated by the prominent Balmer absorption lines 
might result after gas ablation, which in a ram pressure 
stripping event occurs outside-in and reduce the activity of star formation in the outer disc, as indeed observed in local galaxies
\citep[e.g.][]{Boselli+06a, Boselli+16b, Fossati+18}.
We can also mention that the edge-on orientation of the galaxies on the plane of the sky
and in the direction of the stripping favours the detection of the optically thin ablated gas just for projection effects. 

The parallel orientation of the tails of the two galaxies is, however, striking. It is very unlikely that these two objects are members of a group 
and are thus gravitationally interacting while infalling into the cluster. Indeed, their relative velocity is very high ($V_\mathrm{ID473}$ - $V_\mathrm{ID345}$ = 1079 km s$^{-1}$),
thus the duration of the tidal encounter is too short to induce important modifications \citep{Boselli+06}. Furthermore, 
if both galaxies were members of the same infalling group, they should be spiralling around the centre of gravity, producing twisted tails as indeed observed in the
blue infalling group in A1367 \citep{Cortese+06, Fossati+19}.
We remark that both tails are oriented from the galaxies to the south-east, the same direction of the 
filamentary structure joining the cluster CGr32 to the COSMOS Wall \citep{Iovino+16}. This might indicate that both galaxies have been accreted from this structure
and have just crossed the cluster. Their difference in velocity might just indicate that they come from a different region of the COSMOS Wall, which seems extended several Mpc along the line-of-sight
\citep{Iovino+16}.

If the gas keeps the dynamical imprint of the region from where it has been stripped and is decelerated by the interaction with the hot ICM,
as indeed observed in local edge-on galaxy UGC 6697 in A1367 \citep{Consolandi+17}, Fig. \ref{kin} can be interpreted as follows: 
in the northern galaxy ID473 the gas is decelerated in the tail with respect to the galaxy, which means that the trajectory of the galaxy within the cluster is not on the plane of the sky
but it is rather on the line of sight, thus the galaxy is crossing the cluster from the background. This could explain why the tail is relatively short compared to that of ID345. 
Indeed, in the southern galaxy ID345, the less pronounced velocity offset between the galaxy and the tail, and the extension of the tail, suggest 
that the trajectory of the galaxy within the cluster is mainly on the plane of the sky.
The observed large velocities along the tail may be a due to a combination of gas in the tail being decelerated and gas being stripped from the receding side of the galaxy.
The two galaxies show a relatively high velocity dispersion of the gas both in their discs and in the tails ($\sim 100$ km s$^{-1}$). Given their almost edge-on trajectories, this could result 
from the contribution of the gas stripped from both the approaching and receding sides of the rotating disc, whose contribution cannot be resolved because of beam smearing, as indeed observed in 
the edge-on galaxy UGC 6697 in A1367 \citep{Consolandi+17}. We notice, however, that the velocity dispersion of the gas in the tail is smaller than the maximal rotational difference
between the two sides of both galaxies ($\sim$ 150-200 km s$^{-1}$). To conclude, ID345 has probably already passed pericentre, while ID473 is a fresh infaller as suggested by its position in the
phase-space diagram. Galaxies with properties similar to ID345 (eccentric position in the phase-space and tails oriented toward the cluster centre), although unusual, are also present in
local clusters \citep[e.g. NGC 4569,][]{Boselli+14a, Boselli+16a}.

We can also compare the properties of the two galaxies to the most recent results from cosmological hydrodynamic simulations, and in particular to the predictions of IllustrisTNG \citep{Yun+19}
which are perfectly tuned for a comparison with our target since done for galaxies ($M_\mathrm{star}$ $\geq$ 10$^{9.5}$ M$_{\sun}$) and clusters 
(10$^{13}$ M$_{\sun}$ $\leq$ $M_{200}$ $\leq$ 10$^{14.6}$ M$_{\sun}$) of similar mass. These simulations indicate that galaxies with tails of stripped gas due to ram pressure stripping are frequent in clusters with  
a redshift $z<$0.6. About 30\%~ of the disc galaxies have such a cometary shape in the gaseous 
component\footnote{Since these simulations do not discriminate between the different gas phases, it is not astonishing that the fraction of galaxies with observed tails in the ionised gas is 
significantly lower than 30\%.}, with a very week dependence on redshift. 
They also show that these objects are more frequent at intermediate to large
clustercentric radii ($r/R_{200}$ $\geq$ 0.25, that for the cluster CGr32 corresponds to $\simeq$ 250-400 kpc), in galaxies with a high velocity with respect to 
the mean velocity of the cluster, and in more massive clusters. The simulations also show that the orientation of the tail, 
which generally traces the trajectory of the galaxy within the high-density region, does not depend on the relative position of the galaxy
with respect to the cluster centre. We recall that ID345 and ID473 are located at $\simeq$ 200 kpc in projection from the cluster core, have the tails oriented 
almost towards the cluster centre, and have a high velocity along the line of sight with respect to that of the cluster, and thus match
all the properties of perturbed galaxies in the simulations.

\section{Conclusion}
 
As part of the MUSE-gAlaxy Groups In Cosmos (MAGIC) GTO program we observed the 
cluster of galaxies CGr32 ($M_{200}$ $\simeq$ 2 $\times$ 10$^{14}$ M$_{\sun}$) at $z$=0.73. The MUSE observations reveal the presence of two massive ($M_\mathrm{star}$ $\simeq$ 10$^{10}$ M$_{\sun}$) 
galaxies with extended low-surface brightness tails of diffuse gas detected in the [\ion{O}{ii}] emission-line doublet. The cometary shape of the tails and the lack of 
any diffuse stellar counterpart in the very deep Subaru images clearly indicate that the two galaxies are undergoing a ram-pressure stripping event. 
This result is thus the first evidence that the dynamical interaction between the cold galaxy ISM with the hot and dense ICM,
which is probably the dominant perturbing mechanism in high density environments in the local Universe, was also active at earlier epochs, 
when the Universe had only half of its present age. This result is consistent with the predictions of the most recent cosmological simulations of galaxy evolution.

The density of the gas within the tail derived using the [\ion{O}{ii}]$\lambda$3729/[\ion{O}{ii}]$\lambda$3726 line ratio indicates that the recombination time, 
$\tau_\mathrm{rec}$ $\simeq$ 10$^4$ yr, is very short compared to the travel time of the tail, $\tau_\mathrm{travel}$ $\simeq$ 10$^8$ yr, as often observed in local galaxies.
This suggests that a source of gas excitation must be present within the tail.
To conclude, this work is a further confirmation that the extraordinary 
IFU capabilities of MUSE in terms of field of view and sensitivity can be successfully used to extend local studies to the high redshift Universe, thus opening a new era in the study of the role 
of the environment on galaxy evolution.

\begin{acknowledgements}

We thank the anonymous referee for constructive comments and suggestions.
This work was supported by the Programme National Cosmology et Galaxies (PNCG) of CNRS/INSU with INP and IN2P3, co-funded by CEA and CNES.
This work has been carried out thanks to the support of the ANR FOGHAR (ANR-13-BS05-0010-02), the OCEVU Labex (ANR-11-LABX-0060) and the A*MIDEX project (ANR-11-IDEX-0001-02) funded by the ``Investissements d'Avenir'' French government program managed by the ANR. 
VA acknowledges the COLCIENCIAS (Colombia) PhD fellowship program No. 756-2016.
JB acknowledges support by FCT/MCTES through national funds by grant UID/FIS/04434/2019 and through Investigador FCT Contract No. IF/01654/2014/CP1215/CT0003.
This work made use of  observations obtained with XMM-\textit{Newton}, an ESA science mission with instruments and contributions directly funded by ESA Member States and NASA'.

\end{acknowledgements}

\bibliographystyle{aa}
\bibliography{tail_biblio.bib}

\end{document}